\documentclass[prd,twocolumn,preprintnumbers,floatfix,aps,nofootinbib,showpacs]{revtex4-1}

\usepackage{latexsym}
\usepackage{epsfig}
\usepackage{amssymb}

\newcommand{\ba}{\begin{eqnarray}}
\newcommand{\ea}{\end{eqnarray}}
\newcommand{\be}{\begin{equation}}
\newcommand{\ee}{\end{equation}}

\newcommand{\al}{\alpha}

\newcommand{\da}{\delta}

\begin{document}

\title{Bouncing Palatini cosmologies and their perturbations}
\date{\today}

\author{Tomi S. Koivisto}
\email{T.S.Koivisto@uu.nl}
\affiliation{Institute for Theoretical Physics and Spinoza Institute, Postbus 80.195, 3508 TD Utrecht, The Netherlands.}


\begin{abstract}

Nonsingular cosmologies are investigated in the framework of f(R) gravity within the first order formalism.
General conditions for bounces in isotropic and homogeneous cosmology are presented. It is shown that only a quadratic curvature correction is needed to predict a bounce in a flat or to describe cyclic evolution in a curved dust-filled universe. Formalism for perturbations in these models is set up. In the simplest cases, the perturbations diverge at the turnover. Conditions to obtain smooth evolution are derived.

\end{abstract}

\maketitle

\section{Introduction}

Singularities occuring in General Relativity (GR) may be avoided in
more fundamental frameworks where GR predictions are recovered as
the low energy limit. Such frameworks could be provided by String or
M-theory, or loop quantum gravity. In particular, the evolution of the
Big Bang cosmology might be extended to a preceding contracting phase
which, due to new physics relevant at high curvature or energy scales, turns
into the expanding phase our universe is experiencing now \cite{Veneziano:1991ek,Gasperini:1992em}, thus
avoiding the Big Bang singularity (which in the inflationary picture \cite{Starobinsky:1980te}
is manifest rather as geodesic incompleteness \cite{Borde:2001nh} than divergence of curvature invariants).
These scenarios are called bouncing cosmologies \cite{Novello:2008ra}.

Bouncing cosmologies can solve the horizon problem, but to replace inflation
they should, among other things, also predict a viable, nearly scale invariant spectrum of perturbations.
This tricky issue can be circumvented if a curvaton field is responsible for the
generation of fluctuations \cite{Enqvist:2001zp}. Otherwise, matching conditions are
often required to track the evolution of the perturbations across the bounce \cite{Martin:2004pm}.
It has been noted that the curvature perturbation may become singular at the bounce, while
the gravitational potential, whose growing mode usually persists in the post-bounce era,
may have regular behavior \cite{Durrer:2002jn}. A general solution for the perturbations supports
these conclusions \cite{Bozza:2005qg}. It is also well known that the features of the spectrum can
depend sensitively upon the details of the dynamics of the bounce and the physics behind it.

Hence it is useful to consider explicit examples which allow one to scrutinize the possible behaviors of
fluctuations at the bounce. However, it is necessary to violate energy conditions (EC). At least the strong EC
must be broken to change the sign of the expansion rate, and the null EC cannot be respected if there is no
curvature. This rather generically introduces pathologies that, though one may interpret them as only a
shortcoming of the effective theory, hinder from reaching definite conclusions of the evolution of the
spectra \cite{Tsujikawa:2002qc,Martin:2001ue,Kallosh:2007ad}. An example is the perturbation divergence in pre-Big Bang
cosmology \cite{Kawai:1998ab} which can be shown to be an indication of an appearance of a ghost \cite{Koivisto:2006ai}.
To avoid EC violating matter fields, one can contemplate modifications of gravity that
introduce no new degrees of freedom. This can be achieved with an action involving an infinite series of d'Alembertians 
acting on the curvature invariants in such a way that the propagator has no poles;
these string-inspired nonlocal models\footnote{A biscalar-tensor model \cite{Nojiri:2007uq,Koivisto:2008xfa} motivated by the nonlocal cosmology based on inverse d'Alembertian gravity \cite{Wetterich:1997bz,Deser:2007jk}
may also accommodate bounces \cite{Farajollahi:2010pn}, though their viability remains to be shown \cite{Koivisto:2008dh}. Bounces in modified gravity were considered also in e.g. \cite{Aref'eva:2007uk, Saidov:2010wx,Cai:2009in,Abramo:2009qk}.} have been shown to be ghost-free and
asymptotically free at their Newtonian limit \cite{Biswas:2005qr,Biswas:2006bs}.

In the present paper we consider the simpler case of second order
$f(R)$ gravity, which has similar desiderable features. The fourth order metric $f(R)$ models correspond to
scalar tensor theories of the form $\mathcal{L}_{4th} = \phi R + V(\phi)$. By erasing the kinetic term implicit in the
nonminimal coupling, one obtains the second order theory $\mathcal{L}_{2nd} = \phi R -
\frac{3}{4\phi}(\partial\phi)^2 + V(\phi)$. Though this can be problematic in view of the
well-posedness of the Cauchy problem \cite{LanahanTremblay:2007sg}, these simple models
may avoid generic instabilities present in higher order theories \cite{Koivisto:2009jn}
and thus provide an effective description of low energy effects of quantum
gravity. In particular, loop quantum gravity is expected to modify the cosmological
dynamics at high curvatures without introducing new degrees of freedom. To obtain
the quadratic density correction appearing in the particular scalar loop quantum cosmology, one may need to consider
an infinite number of terms in the potential $V(\phi)$, which may be interpreted to reflect
the nonlocal nature of the underlying theory as discussed in Ref.\cite{Olmo:2008nf}. 
Note also the recent extension of the framework \cite{Fatibene:2010yc}.

Low curvature corrections in these so called Palatini-$f(R)$ theories have also been considered as alternatives to dark
energy \cite{Vollick:2003aw,Allemandi:2005qs}, but though they may produce viable background expansion, they generically
fail to produce the observed matter power spectrum, at least for
pressureless dust cosmology \cite{Koivisto:2006ie,Li:2006ag,Uddin:2007gj}.
Problems may appear also at microscopic level, as discussions of electron-electron scattering and Hydrogen atoms, 
seem to imply \cite{Flanagan:2003rb,Olmo:2008ye}, see also \cite{Iglesias:2007nv,Barausse:2007pn}.
This may be due to need to reconsider the averaging problem in these models \cite{Li:2008fa}, or the coupling
of gravity to matter taking torsion and nonmetricity into account \cite{Sotiriou:2006qn,Capozziello:2009mq,Sotiriou:2009xt}.
One may adopt the approach of considering the formalism as an effective macroscopic description, and then
new phenomenology can emerge from the potentially viable high curvature corrections.
Studies of spherically symmetric systems show that the classic Solar system tests are passed
by these models \cite{Kainulainen:2006wz}, while the high curvature effects have interesting predictions for white dwarfs
and neutron stars \cite{Reijonen:2009hi}. Bounces have been suggested too \cite{Barragan:2009sq,Barragan:2009sq}.

We briefly review the first order formalism approach to nonlinear curvature gravity in section \ref{Palatini}
where we also write and solve the cosmological background equations taking into account spatial curvature. We
derive the conditions for bounces to occur and confirm them numerically. In
section \ref{Perturbations} we present the equations governing the evolution of perturbations in convenient forms, based on derivations
 in Refs.\cite{Koivisto:2005yc,Koivisto:2007sq}. Corrections to and generalizations of previous literature are pointed for both the background and the fluctuation equations.
In section \ref{Conclusions} we discuss these results and their implications.

\section{Bouncing backgrounds in Palatini-f(R) gravity}
\label{Palatini}

After writing the general equations for the generalized gravity model, we derive the bouncing conditions and analyze them both analytically and numerically.

\subsection{Palatini approach to generalized gravity}

Consider gravity theories represented by the action
 \be \label{action_p}
 S = \int d^n x \sqrt{-g}
    \left[\frac{1}{2}f(g^{\mu\nu}\hat{R}_{\mu\nu})
    + \mathcal{L}_m(g_{\mu\nu},\phi,...)\right].
 \ee
Here $\phi,...$ are some matter fields. In the Palatini approach one
lets the torsionless connection
$\hat{\Gamma}^\alpha_{\beta\gamma}$ vary independently of the metric.
The Ricci tensor is constructed
solely from this connection,
 \be \label{ricci}
 \hat{R}_{\mu\nu} \equiv \hat{\Gamma}^\alpha_{\mu\nu , \alpha}
       - \hat{\Gamma}^\alpha_{\mu\alpha , \nu}
       +
\hat{\Gamma}^\alpha_{\alpha\lambda}\hat{\Gamma}^\lambda_{\mu\nu}
        -
\hat{\Gamma}^\alpha_{\mu\lambda}\hat{\Gamma}^\lambda_{\alpha\nu}\,.
 \ee
The field equations which follow from extremization of the action
Eq.(\ref{action_p}) with respect to metric variations, can be
written as
 \be \label{fields2} F R^\mu_\nu -\frac{1}{2}f\delta^\mu_\nu =
 T^\mu_\nu \,,
 \ee
where we have defined $F \equiv \partial f/\partial R$.  In GR, $(R-2\Lambda)/8\pi G$, so $F = 1/8\pi G$. By
varying the
action with respect to $\hat{\Gamma}^\alpha_{\beta\gamma}$, obtains
 \be
 \hat{\nabla}_\alpha\left[\sqrt{-g}g^{\beta\gamma}F\right]=0\,,
 \ee
implying that this connection is compatible with the conformal metric
 \be \label{conformal}
 \hat{g}_{\mu\nu} \equiv F^{2/(n-2)}g_{\mu\nu} \,.
 \ee
This connection governs how the tensor $R_{\mu\nu}$ appearing in the
action settles itself, but it turns out that the metric connection
determines the geodesics that freely falling particles follow, since the
energy
momentum
 \be \label{memt}
 T_{\mu\nu} \equiv -\frac{2}{\sqrt{-g}} \frac{\delta
 (\sqrt{-g}\mathcal{L}_m)}{\delta(g^{\mu\nu})}\,.
 \ee
is conserved according to this connection,
 \be \label{em}
 \nabla_\mu T^\mu_\nu = 0\,,
 \ee
whereas in general $\hat{\nabla}_\mu T^\mu_\nu \neq 0$. Therefore we have a metric theory of gravity \cite{Koivisto:2005yk}
in the sense of Ref.\cite{Will:2005va}.
The trace of the field equations allows
us to solve $R$ as an algebraic function of the matter trace $T \equiv
g^{\mu\nu}T_{\mu\nu}$. This central relation reads
 \be \label{trace} FR-2f = T\,.
 \ee
From now on we set the spacetime dimension to $n=4$ and use units $8 \pi
G \equiv c \equiv 1$. Written in the form of
GR plus correction terms, the field equations read:
 \ba \label{eg} G^\mu_\nu(g) & = & {T^\mu_\nu}
               + (1-F)R^\mu_\nu(g)
               -  \frac{3}{2F}(\nabla^\mu F)(\nabla_\nu F) \nonumber \\
\nonumber
               +   \nabla^\mu \nabla_\nu F
             &  + & \frac{1}{2}\left[(f-R) + (1-\frac{3}{F})\Box F +
     \frac{3}{2F}(\partial F)^2\right]\delta^\mu_\nu.
 \ea
Since the corrections can be expressed as functions of the matter trace,
one can view Eq.(\ref{eg}) as GR with generalised coupling to matter: only the way that
''matter tells spacetime how to curve'' is modified. So, the whole RHS may be
regarded as an effective matter energy-momentum tensor.
In vacuum it reduces to a cosmological constant \cite{Ferraris:1992dx}.
This is also the case in the presence of conformal
matter, i.e. if $T=0$.

\subsection{Background cosmology}

In the spatially flat Friedmann-Lema\'itre-Robertson-Walker (FLRW)
universe with the line element
\be
ds^2 = -dt^2 + a^2(t)\left(\frac{dr^2}{1-Kr^2}+r^2d\Sigma^2\right)\,,
\ee
and a perfect fluid source with a constant equation of state $w=p/\rho$, the
Friedmann equation can be written as
\be \label{fried}
6F\left(H+\frac{\dot{F}}{2FH}\right)^2 = \rho+3p + f -
6F\frac{K}{a^2}\,.
\ee
In the Appendix \ref{hlaw} we write the general Friedmann
equation when $\dot{w}=0$ is not assumed.
The Hubble parameter can be expressed fully in terms of the curvature
scalar, and in our case then rewritten as
\be \label
6H^2 = \label{hoo}
\frac{1}{(1-3w)F}\frac{3(1+w)f-(1+3w)FR-6F\frac{K}{a^2}}{\left(1-\frac{3}{2}(1+w)\frac{F'(RF-2f)}{F(RF'-F)}\right)^2} \,.
\ee
We have expressed the Hubble rate as a function of $R$, which we in turn
may solve from the trace equation (\ref{trace}). If the scale factor is monotonic, one may
find its evolution algebraically once the given matter content as outlined
in Ref. \cite{Amarzguioui:2005zq}. In our case however it is preferable
to solve the numerical system by the integrating differential equations, which are shown in the
appendix.

\subsection{Bouncing solutions}
\label{quad}

In general, a necessary condition for a bounce to occur is obtained from (\ref{hoo}) and (\ref{trace}) as
\ba \label{type1}
F& = & 0\,, \quad \text{or}  \\
F(12\frac{K}{a^2}-R)& = & 3\rho(1+w)\,. \label{type2a}
\ea
Now, by solving the trace equation (\ref{trace}),  $FR$ is given by an inverse function of the density, and its form is very dependent of $f(R)$.
The quadratic term, which may be considered the leading correction\footnote{In addition, one notes that this is the only case when the trace equation (\ref{trace}) is linear in the sources. One may then contemplate if more general functions $f(R)$ would be, after averaging, effectively described by the quadratic model.}
to GR, can already lead to bouncing cosmology.
Let us thus consider the case
\be \label{quadratic}
f(R) = R + \alpha R^2\,,
\ee
to study explicitly the background behavior. This model results in a symmetric bounce (the
pre-Big Bang is the time reversal of the post-Big Bang evolution).
The Friedmann equation is now
\be \label{qh}
3H^2=\frac{a^3+2\alpha R_0}{2a^3(a^3-\al R_0)^2}\left[2a^3R_0-6Ka(a^3+2\al R_0)+\al R_0^2\right]\,,
\ee
where $R_0>0$ is a constant which is equal to the matter density at $a=1$. The derivatives of the Hubble rate
are written in the appendix as (\ref{qh1}) and (\ref{qh2}).
The bounce condition (\ref{type1},\ref{type2a}) now becomes
\ba
a^3+2\al R_0 & = & 0\,, \quad \text{or} \nonumber \\
(2a^3+\al R_0)R_0 & = & 6K a(a^3+2\al R_0)\,.
\ea
In the flat case, the second condition is simply $a^3=-\alpha R_0/2$, which can be satisfied given a negative $\al<0$.
However, the first condition will be saturated earlier when the scale factor has contracted to $a^3=-2\alpha R_0$.
Then, at the bounce we have $F\rightarrow 0$.

Let us consider whether one may avoid $F=0$ at the turnover in curved models with $K \neq 0$.
Since the normalization of the scale factor is arbitrary, let us assume here that the second bounce
condition is fulfilled before the first one at $a=1$. This implies that
\be
2\alpha R_0 > -1 \,,
\ee
and that
\be
R_0 = 6K\frac{1+2\alpha R_0}{2+\al R_0}\,.
\ee
Solving $\alpha$ from the second constraint and plugging into the first condition gives
\be
\frac{4(3K-R_0)}{R_0-12K} > -1\,.
\ee
We should assume $R_0>12K$, since in any realistic universe the curvature is subdominant to the matter density
at early times by many orders of magnitude. Then the constraint reduces to $R_0<0$, in contradiction to our assumptions.
If we however allow a positive curvature to dominate over matter density at the bounce, $K>R_0/12$, we can realize bounces where
$F$ stays finite at the turning point. This could have occurred if there was significant amount of inflation after the bounce which diluted
away both the curvature and matter density. However, the details of such a case are not of interest to us, as there also the possible
signatures from bounce were most probably erased.

As we will see in the following, the $F=0$ bounce results in divergence of perturbations. In general, assuming negligible curvature
but allowing general $f(R)$ and $w$, the condition for the second type of bounce (\ref{type2a}) assumes the very simple form
\be \label{type2}
F>0\,, \quad f(R)=-\rho(1+3w)\,.
\ee
In the simplest models considered here in detail, this is a necessary condition for perturbations to stay regular at the bounce. In section \ref{Conclusions} we briefly discuss possibly viable generalizations of the models.

\begin{figure}
\includegraphics[width=7.5cm]{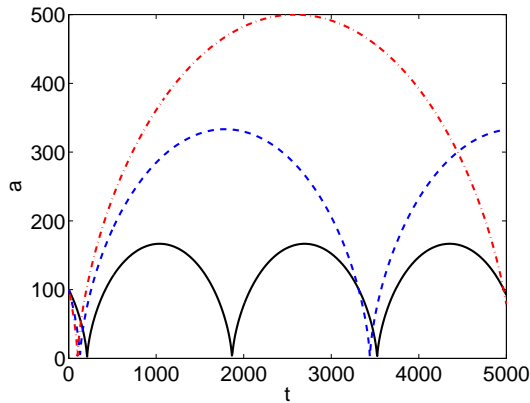}
\caption{\label{a_fig} Solid (black) line: cyclic evolution of the scale factor in the quadratic model with $K>0$. Dashed
(blue) line: the same model with doubly as much matter. Dash-dotted (red) line: the same model with tripled matter density.}
\end{figure}

\section{Perturbations}
\label{Perturbations}

In the following we first specify our perturbation system, then consider the evolution of
the perturbations in pressureless matter as a specific example, and its description in terms
of the canonical variable. Cosmological perturbation theory is presented in the reviews \cite{Kodama:1985bj,Mukhanov:1990me},
and applied to this class of generalized gravity theories in Refs. \cite{Koivisto:2005yc,Koivisto:2007sq}.

\subsection{The perturbation system}

The line-element in the perturbed Friedmann-Lemaitre-Robertson-Walker
(FLRW) spacetime can be written as
 \ba \label{metric}
  ds^2 & = & a^2(t)\Big\{-\left(1+2\phi\right)dt^2 + b_i dx^idt
\nonumber \\
       & + & \left[ g^{(3)}_{ij}+2\left(g^{(3)}_{ij}\psi + h_{ij}\right)
\right]dx^idx^j\Big\}.
 \ea
We characterize the scalar perturbations in the longitudinal Newtonian
gauge by the variables gravitational potentials $\phi$ and $\psi$.
Vector perturbations introduce two more degrees of freedom, encoded here
into the divergenceless 3-vector field $b_i$.
Gravitational waves are described by the two free components of the
symmetric, transverse and traceless 3-tensor $h_{ij}$.
The comoving spatial background metric $g^{(3)}_{ij}$ reduces to
$\delta_{ij}$ in a flat universe.
The vertical bar indicates a covariant derivative based on the
Levi-Civita connection of $g^{(3)}_{ij}$. This metric is used to lower
and raise spatial indices $i,j,k \dots$ of the perturbation variables.
The components of the energy-momentum tensor for a general fluid is
imperfect fluid
are
 \ba \label{fluid}
 T^0_0 & = & -(\bar{\rho}+\delta\rho)\,, \\
 T^0_i & = & -\left(\bar{\rho}+\bar{p}\right)\left(v{,_i} +
 v^{(v)}_i\right)\;, \\ T^i_j & = & (\bar{p}+\delta p)\delta^i_j +
 \Pi^i_j\,.
 \ea
Here $\rho$ and $p$ are energy density and pressure, and $v$, $v^{(v)}$
are the scalar and vector velocity perturbations, respectively.
Background quantities are denoted with an overbar, which we will usually
omit when unnecessary.
The isotropy of the background does not allow anisotropic stress except
as a perturbation. This we decompose into the scalar, vector and tensor
contributions as
 \be \Pi_{ij} \equiv \left(\Pi^{(s)}_{|ij} +
\frac{1}{3}\triangle\Pi^{(s)} \right) + \Pi^{(v)}_{(i|j)} +
\Pi^{(t)}_{ij}\,, \ee
where $\triangle$ stands for the three-space Laplacian based on the
Levi-Civita connection of $g^{(3)}_{ij}$. The vector $\Pi^{(v)}_i$ is
divergence-free and the tensor $\Pi^{(t)}_{ij}$ is symmetric,
transverse, and traceless. This completes our specification of the
perturbation system.

\subsection{Fluid quantities}

Next we will discuss the evolution of the system in terms of fluid variables. First we
consider the density perturbation in the comoving gauge (i.e. CDM rest
frame in the present case) and
then the velocity perturbation in the uniform-density gauge (i.e. the
frame where CDM is smoothly distributed). The former quantity becomes
ill-defined at the bounce, the latter behaves somewhat better.

It is convenient to introduce the comoving density perturbation $\Delta$ which
is given by longitudinal gauge quantities as follows:
\be \label{com}
\Delta = \delta + 3H(1+w)\frac{av}{k}\,,
\ee
The evolution equation for the perturbations has been derived in the
general case and is of the form
\be \label{s_dddelta}
\ddot{\Delta} = D_1 H\dot{\Delta} + \left(D_2 H^2 + D_k
\frac{k^2}{a^2}\right)\Delta + P_1 H\dot{\Pi} + P_2 H^2\Pi\,,
\ee
where the dimensionless coefficients are given in the appendix of Ref.
\cite{Koivisto:2007sq}.
In the case of pressureless dust the evolution equation simplifies to
\be \label{com_e}
\ddot{\Delta} + \left(2H + \mathcal{F}\right)\dot{\Delta} =
\left(\frac{\ddot{H}}{H} + 2\dot{H} + \frac{\dot{H}}{H}\mathcal{F} -
\frac{k^2}{a^2}c_{eff}^2\right)\Delta\,.
\ee
where we have defined the auxiliary quantity
\be
\mathcal{F} \equiv
\frac{2}{\dot{F}+2FH}\left(\ddot{F}-\frac{\dot{F}^2}{F} -
\frac{\dot{F}\dot{H}}{H}\right)\,.
\ee
and the effective sound speed squared
\be
c_{eff}^2 \equiv \frac{\dot{F}}{3(\dot{F}+2FH)}\,.
\ee
If both $\mathcal{F}$ and $c_{eff}^2$ vanish GR evolution is recovered,
so these variables represent the modified gravity effects.
In the specific example model discussed in section \ref{quad}, both of these terms apparently diverge at the bounce.
In particular, we have divisions by $H$, $\dot{F}+2FH$ and $F$, where the first
term vanishes always, the second term vanishes at least for dust, and the last
term vanishes at least for the flat dust bounce model. In particular, for the quadratic model
\be
c^2_S = -\frac{\alpha R_0}{a^3-\al R_0}\,,
\ee
\be
\mathcal{F} = \frac{18H\alpha R_0 a^3}{(a^3-\alpha R_0)(a^3 + 2\alpha R_0)}\,,
\ee
and we see that when $a^3=-\al R_0$, the sound speed in fact is regular
but the factor $\mathcal{F}$ is not. The $\dot{H}$ and $\ddot{H}/H$ are given in the appendix as (\ref{qh1}) and (\ref{qh2}).
Though the comoving gauge is where we want our observable density in, this coordinate system can
become ill-defined at the bounce. However, a possibility remains that
this is not a physical problem.

The perturbations can be carried across the bounce in another variable
than the comoving gauge density perturbation.
Another convenient quantity to consider is $v_\delta$, the velocity
perturbation of matter evaluated in the uniform-density gauge. From
Eq.(\ref{com}), we have
\be \label{den}
v_\delta=\frac{k}{3aH}\Delta\,,
\ee
and readily obtain from (\ref{s_dddelta}) the evolution equation
which coincides with the Eq.(46) in Ref.\cite{Koivisto:2005yc},
\ba \label{den_e}
\ddot{v}_\da & + & \left( 4H+2\frac{\dot{H}}{H} +
\mathcal{F}\right)\dot{v}_\da \nonumber \\ & + &  \left(3(\dot{H}+H^2) +
H\mathcal{F} + \frac{k^2}{a^2}c_{eff}^2 \right)v_\da\ = 0,.
\ea
The divergent $1/H$ terms in the second expression cancel,
\ba
4H & + & 2\frac{\dot{H}}{H} + \mathcal{F} \nonumber \\
& = &
\frac{2}{\dot{F}+2FH}\left[\ddot{F}-\frac{\dot{F}^2}{F}+2\left(FH\right)^\bullet
+ 4FH^2\right]\,.
\ea
However, the $\dot{F}/F$ can divergent, when $F \rightarrow 0$ in the
flat bounce model. This can be avoided with curvature $K>\rho/12$, or in the bounce of the type (\ref{type2}) but even
then typically $\dot{F}\rightarrow 0$ at the turnover.
Consequently, the factor $\mathcal{F}$ remains apparently divergent due to
$\mathcal{F} \sim \frac{1}{\dot{F}+2FH}$. Thus, by
considering (\ref{den}) instead of (\ref{com}) the apparent divergencies
of the coefficients in the evolution equations can be made less severe (from $\sim 1/H^2$ to $\sim H$) but
they seem to persist.


\subsection{Canonical variable}

The canonical variable $\nu$ obeys the equation of motion
\be \label{can_e}
\ddot{\nu} + H\dot{\nu} + \left(-\frac{3}{4}H^2+\frac{1}{2}\dot{H}-\frac{K}{a^2} + C_\nu +
\frac{k^2-3K}{a^2} c_{eff}^2 \right)\nu = 0\,,
\ee
where
\be \label{cnu}
C_\nu = -\frac{\dot{F}}{2F}\left(3H+\frac{\dot{F}}{F}\right)\,.
\ee
The coefficients remain apparently divergent for the flat dust bounce model of section \ref{quad}, because the potential
becomes infinite due to the $\sim \dot{F}^2/F^2$ term.
However, for the curvature-dominated model or bounce of the type (\ref{type2}) both $C_\nu$ and the sound speed remain regular at the bounce,
since $H,\dot{F} \rightarrow 0$.
Thus we have found the explicit conditions for the perturbations to be carried smoothly across the bounce.
The relation between the Mukhanov-Sasaki variable
$\nu$ and the comoving density perturbation is given by
\be \label{transf}
\nu(t) = \exp\left\{\frac{1}{2}\int^t\left[H(t')+\mathcal{F}(t')\right]dt'\right\}\delta(t)\,.
\ee
Hence, for a model with $F$ not dipping to zero, one may use the regular equation (\ref{can_e}) to solve the
evolution, and the transformation (\ref{transf}) to obtain the results in terms of the observables.


\section{Conclusions}
\label{Conclusions}

Within the Palatini framework one may consider extensions of GR without introducing new degrees of
freedom. Hence they may serve as useful toy models describing more completely the cosmological evolution,
with motivations from e.g. loop quantum cosmology. We showed that there are nonsingular bouncing backgrounds in simple examples of
such quantum corrected gravity models, and set up the formalism for the perturbations in these models in order to
monitor the evolution of their spectra across a bounce. 

The models of the type (\ref{type1}), characterized by $F\rightarrow 0$ at the bounce, were found to feature singular behavior of perturbations in
a flat, dust-filled universe. This may be cured in the curvature-dominated case,
reflecting the fact that in the $K=0$ case the effective matter
sources necessarily violate the null EC, whereas this
is not the case if curvature is present. Since $F$ gives the
sign of the graviton action, pathologies were to be expected at $F \rightarrow 0$. At this point the conformal relation between Einstein and Jordan frames (\ref{conformal}) becomes ill-defined.
Let us note though that as the perturbations explode, their backreaction will render the perturbative system invalid,
 and as these nonlinear effects are very difficult to tackle in practice, we cannot say if there is a true singularity or not and whether the bounce occurs or not. In this light the problem is rather unpredictivity.

Obviously, it would be interesting to explore possible ways to obtain bounces of the type (\ref{type2}).
Such might be constructed by considering just more general functions $f(R)$ than the one involving solely a
monomial correction.

Another way to obtain smooth evolution could be to include more general sources than completely pressureless fluids. Apart from allowing $w\neq 0$,
one may also consider stabilizing the system with entropic or anisotropic stresses.
Indeed, this has previously proved successful in eliminating instabilities in matter perturbations in some dark energy models \cite{Koivisto:2007sq,Koivisto:2005mm} (however, those models
based on infrared gravity modifications may be otherwise problematical as discussed in the introduction). More realistically,
also radiation would be included as a source and this changes the dynamics and possibly the conclusions.
Our results can be directly applied to such more general models
with possibly regular evolution. This is left for future studies.


As a concluding remark we note it is possible that a complete evolution of the background and structures of the universe is not amenable to classical description by second order differential equations, and one may have to take into account in a more nontrivial way the presently unknown physics at the very high curvature scales in order to provide a fully consistent coarse-grained picture of the cosmology that emerges. Meanwhile, the quest for the effective field equations for gravitational interactions, at both high and low curvature regimes, is ongoing.

\appendix

\section{Friedmann equations}
\label{hlaw}

\subsection{General case}

Without assuming a constant equation of state or vanishing curvature, the Hubble parameter may be written as
\ba
H =
\qquad\qquad
\qquad\qquad\qquad\qquad
\qquad\qquad\qquad\qquad
\\ \nonumber  \frac{ 9 \dot{w} F' +
     (F - F' R) \sqrt{3F\left(F (R-12\frac{K}{a^2}) + 3 \rho (1 + w)\right)}}{3 \left(2 F^2 -
     2 F F' R - 3 F' \rho \alpha_w\right)}\,.
\ea
where $\alpha_w=(1+w)(1-3w)$ and $F'=dF/dR$.
For a constant equation of state $\dot{w}=0$, this reduces to
\be
3H^2=\frac{F\left[3\rho(1+w)-F(12\frac{K}{a^2}-R)\right]}{4\left[F+\frac{3}{2}\frac{F'\rho(1+w)(1-3w)}{RF'-F}\right]^2}\,.
\ee
In the limit $K=0$, this agrees with the formulas in the references \cite{Allemandi:2005qs,Koivisto:2005yc,Amarzguioui:2005zq}
(and in several later references). 
However, this does not reduce to the Hubble law considered in Ref.\cite{Barragan:2009sq}, and consequently
our solutions and results in subsection \ref{quad} are somewhat different from theirs. For example, their Eq.(20) for the specific case of the quadratic model is quite different from our Eq.(\ref{qh}) also when $K=0$. Note that in Ref.\cite{Barragan:2010uj} the authors have corrected a mistake regarding nonzero spatial curvature in Ref.\cite{Barragan:2009sq}, and there the considerations are in accordance with ours here.


\subsection{Quadratic model}

The first two derivatives of the Hubble rate in the quadratic model (\ref{quadratic}) are
\begin{widetext}
\be \label{qh1}
\dot{H} = -\frac{
a^7 R_0 \left(a^2-24   \al K\right)-2 a^{10}K
+6 a^4 \al R_0^2 \left(a^2-6 \al K\right)
+a \al^2 R_0^3 \left(3 a^2+8 \al K\right)
-\al^3   R_0^4}{2 \left(a^4-a \al R_0\right)^3}\,.
\ee
\ba \label{qh2}
\frac{\ddot{H}}{H} & = &
\frac{1}{2 \left(a^4-a \al   R_0\right)^4}\Big[a^{10} R_0 \left(3a^2-134 \al K\right) -  4 a^{13}K \nonumber \\
& + & 6 a^7 \al R_0^2 \left(7 a^2-64 \al K\right)
+a^4\al^2 R_0^3 \left(45 a^2+52 \al K\right)
-4 a \al^3 R_0^4 \left(3 a^2+4 \al K\right) + 3 \al^4 R_0^5\Big]\,.
\ea
\end{widetext}
Eq.(\ref{qh1}) is a suitable form for numerical integration. We checked that the solution gives Eq.(\ref{qh}) and that
its numerical derivative reproduces Eq.(\ref{qh2}). The latter algebraic expression is needed specifically at the bounce where
$\ddot{H},H\rightarrow 0$.

\acknowledgments

The author is grateful to T. Biswas and A. Mazumdar for enlightening discussions and to G. Olmo, P. Peter and T. Sotiriou for useful comments on the manuscript.
This work was supported by the FOM and the Finnish Academy.

\appendix

\bibliography{palarefs}

\end{document}